\font\tenfrakturb=eufb10
\font\tenfraktur=eufm10
\font\tenmsbm=msbm10
\font\sevenfrakturb=eufb7
\font\sevenfraktur=eufm7
\font\sevenmsbm=msbm7
\font\fivefrakturb=eufb5
\font\fivefraktur=eufm5
\font\fivemsbm=msbm5
\newfam\bgothicfam
\newfam\gothicfam
\newfam\msbmfam
\textfont\bgothicfam = \tenfrakturb \scriptfont\bgothicfam=\sevenfrakturb
\scriptscriptfont\bgothicfam=\fivefrakturb
\textfont\gothicfam = \tenfraktur \scriptfont\gothicfam=\sevenfraktur
\scriptscriptfont\gothicfam=\fivefraktur
\textfont\msbmfam = \tenmsbm \scriptfont\msbmfam=\sevenmsbm
\scriptscriptfont\msbmfam=\fivemsbm

\def\Bbb{\tenmsbm\fam\msbmfam}

\catcode`@=11
\def\renewcounter#1{\@definecounter{#1}\@ifnextchar[{\@newctr{#1}}{}}

\documentstyle{elsart}
\begin{document}
\begin{frontmatter}
\title{Estimates of the gluon concentrations in the confining
SU(3)-Yang-Mills field for the first three states of charmonium }
\author{Yu. P. Goncharov}
\address{Theoretical Group, Experimental Physics Department, State Polytechnical
         University, Sankt-Petersburg 195251, Russia}
\author{A. A. Bytsenko}
\address{Departamento de Fisica, Universidade Estadual de Londrina,
Caixa Postal 6001, Londrina-Parana, Brazil}

\begin{abstract} 
The estimates of the gluon concentrations in the classical SU(3)-Yang-Mills 
field modelling confinement are given for the first three states of charmonium
whose spectrum is tuned by calculating electromagnetic transitions among the 
mentioned levels in dipole approximation. For comparison the corresponding 
estimates for the photon concentration in the ground state of positronium 
(parapositronium and orthopositronium) are also adduced.
\end{abstract}
\end{frontmatter}

\section{Introduction and preliminary remarks}

As was discussed in Refs. \cite{{Gon01},{GC03},{Gon03}}, the natural way of 
building meson spectroscopy might be based on
the exact solutions of the SU(3)-Yang-Mills equations modelling
quark confinement, the so-called confining solutions. Such solutions are
supposed to contain only the components of the SU(3)-field which are 
Coulomb-like or linear in $r$, the distance between quarks. In 
Ref. \cite{Gon01} a number of such solutions have been obtained and the 
corresponding spectrum of Dirac equation describing the relativistic bound
states in those confining SU(3)-Yang-Mills fields has been analysed. It should
be noted that the given approach is the direct
consequence of the {\it relativistic} QCD (quantum chromodynamics) Lagrangian 
since the mentioned Yang-Mills and Dirac equations are derived just from the 
latter one.

Further in Refs. \cite{{GC03},{Gon03}} the results obtained were successfully 
applied to the description of the quarkonia spectra (charmonium and 
bottomonium). In its turn, the mentioned description points out the 
linear confinement to be (classically) governed by the colour magnetic field 
linear in $r$. 

 The results of Refs. \cite{{Gon01},{GC03},{Gon03}} suggest the following mechanism of 
confinement to occur within the framework of QCD (at any rate, for mesons
and quarkonia). The gluon exchange between quarks is realized in such a way
that at large distances it leads to the confining SU(3)-field which may be 
considered classically (the gluon concentration becomes huge and gluons form 
the boson condensate -- a classical field) and is 
a {\em nonperturbative} solution of the SU(3)-Yang-Mills 
equations. Under the circumstances mesons are the relativistic bound states 
described by the corresponding wave functions -- {\em nonperturbative} 
solutions of the Dirac equation in this 
confining SU(3)-field \cite{{Gon01},{GC03},{Gon03}}. For each meson there 
exists its own set of real constants (for more details see below) 
$a_j, A_j, b_j, B_j$ parametrizing the 
mentioned confining gluon
field (the gluon condensate) and the corresponding wave 
functions while the latter ones also depend on $\mu_0$, the reduced
mass of the current masses of quarks forming 
meson \cite{{Gon01},{GC03},{Gon03}}. It is clear that constants 
$a_j, A_j, b_j, B_j,\mu_0$
should be extracted from experimental data and such a program has been just 
realized in Refs. \cite{{GC03},{Gon03}} for quarkonia.

The aim of the present paper is to estimate the above gluon concentrations
for the case of charmonium but we shall restrict ourselves to 
the first three states of charmonium. For comparison we shall
also adduce the corresponding estimates for the
photon concentrations in the ground state of positronium (parapositronium and
orthopositronium) to emphasize some analogy between QCD and QED (quantum
electrodynamics).

Further we shall deal with the metric of
the flat Minkowski spacetime $M$ that
we write down (using the ordinary set of local spherical coordinates
$r,\vartheta,\varphi$ for the spatial part) in the form
$$ds^2=g_{\mu\nu}dx^\mu\otimes dx^\nu\equiv
dt^2-dr^2-r^2(d\vartheta^2+\sin^2\vartheta d\varphi^2)\>. \eqno(1)$$
Besides, we have $|\delta|=|\det(g_{\mu\nu})|=(r^2\sin\vartheta)^2$
and $0\leq r<\infty$, $0\leq\vartheta<\pi$,
$0\leq\varphi<2\pi$.

Throughout the paper we employ the system of units with $\hbar=c=1$,
unless explicitly stated otherwise.
Finally, we shall denote $L_2(F)$ the set of the modulo square integrable
complex functions on any manifold $F$ furnished with an integration measure
while $L^n_2(F)$ will be the $n$-fold direct product of $L_2(F)$
endowed with the obvious scalar product.

If $A=A_\mu dx^\mu=A^c_\mu \lambda_cdx^\mu$ is a SU(3)-connection in the
(trivial) three-dimensional bundle $\xi$ over the Minkowski spacetime, 
where $\lambda_c$ are the known Gell-Mann matrices, then
we are interested in the confining solutions $A$ of the SU(3)-Yang-Mills 
equations
$$d\ast F= g(\ast F\wedge A - A\wedge\ast F) \>\eqno(2)$$ 
with the exterior differential $d=\partial_t dt+\partial_r dr+
\partial_\vartheta d\vartheta+\partial_\varphi d\varphi$ in coordinates
$t,r,\vartheta,\varphi$,
while the curvature matrix (field strentgh)
for the $\xi$-bundle is 
$F=dA+gA\wedge A= F^a_{\mu\nu}\lambda_adx^\mu\wedge dx^\nu$ and $\ast$ 
means the Hodge star
operator conforming to metric (1), $g$ is a gauge coupling constant.

For the aims of the given paper we shall use the confining solution of 
Ref. \cite{Gon01} in the form
$$ A^3_t+\frac{1}{\sqrt{3}}A^8_t =-\frac{a_1}{r}+A_1 \>,
 -A^3_t+\frac{1}{\sqrt{3}}A^8_t=\frac{a_1+a_2}{r}-(A_1+A_2)\>,
-\frac{2}{\sqrt{3}}A^8_t=-\frac{a_2}{r}+A_2\>, $$
$$ A^3_\varphi+\frac{1}{\sqrt{3}}A^8_\varphi =b_1r+B_1 \>,
 -A^3_\varphi+\frac{1}{\sqrt{3}}A^8_\varphi=-(b_1+b_2)r-(B_1+B_2)\>,
-\frac{2}{\sqrt{3}}A^8_\varphi=b_2r+B_2\> \eqno(3)$$
with all other $A^c_\mu=0$, where real constants $a_j, A_j, b_j, B_j$
parametrize the solution, and we wrote down
the solution in the combinations that are just
needed to insert into the Dirac equation
$${\cal D}\Psi=\mu_0\Psi\>\eqno(4)$$
with the relativistic wave function $\Psi=(\Psi_1, \Psi_2, \Psi_3)$ 
of the quarkonium,
where the four-dimensional spinors $\Psi_j$ represent the $j$-th colour
component of the quarkonium, 
$\mu_0$ is a mass parameter and one can consider it to be the reduced
relativistic mass which is equal, {\it e. g.} for quarkonia, to half the 
current mass of quarks forming a quarkonium,
while the coordinate $r$ stands for the distance between quarks.
The explicit form of operator ${\cal D}$ is not needed here and
can be found in
Refs. \cite{{Gon01},{GC03},{Gon03}}.

Finally, for the necessary estimates we shall employ the $T_{00}$-component of 
the energy-momentum tensor for a SU(3)-Yang-Mills field
$$T_{\mu\nu}={1\over4\pi}\left(-F^a_{\mu\alpha}\,F^a_{\nu\beta}\,g^{\alpha\beta}+
{1\over4}F^a_{\beta\gamma}\,F^a_{\alpha\delta}g^{\alpha\beta}g^{\gamma\delta}
g_{\mu\nu}\right)\>. \eqno(5) $$
\section{Relativistic spectrum of charmonium}
From the adduced form it is clear that the solution (3) 
is a configuration describing the electric Coulomb-like colour field 
(components $A_t$) and the magnetic colour field linear in $r$ (components 
$A_\varphi$) so the solution can be rewritten as
$$A^3_t = -(a_1+a_2/2)/r+A_1+A_2/2,\>
A^8_t =(a_2/r-A_2)\sqrt{3}/2\>,$$
$$ A^3_\varphi = (b_1+b_2/2)r+B_1+B_2/2,
   A^8_\varphi= -(b_2r+B_2)\sqrt{3}/2\>.\eqno(6)$$
 As was shown in Ref. \cite{Gon01}, after inserting the above confining
solution into Eq. (4), it admits the solutions of the form (with Pauli 
matrix $\sigma_1$)
$$\Psi_j=e^{i\omega_j t}r^{-1}\pmatrix{F_{j1}(r)\Phi_j(\vartheta,\varphi)\cr\
F_{j2}(r)\sigma_1\Phi_j(\vartheta,\varphi)}\>,j=1,2,3\eqno(7)$$
with the 2D eigenspinor $\Phi_j=\pmatrix{\Phi_{j1}\cr\Phi_{j2}}$ of the
euclidean Dirac operator on the unit sphere ${\Bbb S}^2$.
The general explicit form of $\Phi_j$ is not needed here and
can be found in
Refs. \cite{Gon99}. For the purpose of the present paper we shall adduce 
the necessary spinors below. Spinors $\Phi_j$ form an orthonormal basis in 
$L_2^2({\Bbb S}^2)$ and they can be subject to
the normalization condition
$$\int_{{\Bbb S}^2}\Phi^{\dag}_{j}\Phi_{j}d\Omega=
\int\limits_0^\pi\,\int\limits_0^{2\pi}(|\Phi_{j1}|^2+|\Phi_{j2}|^2)
\sin\vartheta d\vartheta d\varphi=1\> , \eqno(8)$$
where ($\dag$) stands for hermitian conjugation.

The energy spectrum $\varepsilon$ of quarkonium is given by the
relation $\varepsilon=\omega_1+\omega_2+\omega_3$ with
$$\omega_1=\omega_1(n_1,l_1,\lambda_1)= gA_1+
\frac{-\Lambda_1 g^2a_1b_1+(n_1+\alpha_1)
\sqrt{(n_1^2+2n_1\alpha_1+\Lambda_1^2)\mu_0^2+g^2b_1^2(n_1^2+2n_1\alpha_1)}}
{n_1^2+2n_1\alpha_1+\Lambda_1^2}\>,\eqno(9)$$
$$\omega_2=\omega_2(n_2,l_2,\lambda_2)= -g(A_1+A_2)+$$
$$\frac{-\Lambda_2 g^2(a_1+a_2)(b_1+b_2)-(n_2+\alpha_2)
\sqrt{(n_2^2+2n_2\alpha_2+\Lambda_2^2)\mu_0^2+g^2(b_1+b_2)^2
(n_2^2+2n_2\alpha_2)}}
{n_2^2+2n_2\alpha_2+\Lambda_2^2}\>,\eqno(10)$$
$$\omega_3=\omega_3(n_3,l_3,\lambda_3)=gA_2+
\frac{-\Lambda_3 g^2a_2b_2+(n_3+\alpha_3)
\sqrt{(n_3^2+2n_3\alpha_3+\Lambda_3^2)\mu_0^2+g^2b_2^2(n_3^2+2n_3\alpha_3)}}
{n_3^2+2n_3\alpha_3+\Lambda_3^2}\>,\eqno(11)$$
where
$\Lambda_1=\lambda_1-gB_1\>,\Lambda_2=\lambda_2+g(B_1+B_2)\>,
\Lambda_3=\lambda_3-gB_2\>,$
$n_j=0,1,2,...$, while $\lambda_j=\pm(l_j+1)$ are
the eigenvalues of euclidean Dirac operator
on unit sphere with $l_j=0,1,2,...$ Besides
$$\alpha_1=\sqrt{\Lambda_1^2-g^2a_1^2}\>,
\alpha_2=\sqrt{\Lambda_2^2-g^2(a_1+a_2)^2}\>,
\alpha_3=\sqrt{\Lambda_3^2-g^2a_2^2}\>.\eqno(12)$$

Further, the radial parts of (7) are given at $n_j=0$ by
$$F_{j1}=C_jP_jr^{\alpha_j}e^{-\beta_jr}\left(1-
\frac{Y_j}{Z_j}\right),F_{j2}=iC_jQ_jr^{\alpha_j}e^{-\beta_jr}\left(1+
\frac{Y_j}{Z_j}\right),\eqno(13)$$
while at $n_j>0$ by
$$F_{j1}=C_jP_jr^{\alpha_j}e^{-\beta_jr}\left[\left(1-
\frac{Y_j}{Z_j}\right)L^{2\alpha_j}_{n_j}(r_j)+
\frac{P_jQ_j}{Z_j}r_jL^{2\alpha_j+1}_{n_j-1}(r_j)\right],$$
$$F_{j2}=iC_jQ_jr^{\alpha_j}e^{-\beta_jr}\left[\left(1+
\frac{Y_j}{Z_j}\right)L^{2\alpha_j}_{n_j}(r_j)-
\frac{P_jQ_j}{Z_j}r_jL^{2\alpha_j+1}_{n_j-1}(r_j)\right],\eqno(14)$$
with the Laguerre polynomials $L^\rho_{n_j}(r_j)$, $r_j=2\beta_jr$,
where, for $j=1$, $\beta_1=\sqrt{\mu_0^2-(\omega_1-gA_1)^2+g^2b_1^2}$,
$P_1=gb_1+\beta_1$, $Q_1=\mu_0+\omega_1-gA_1$,
$Y_1=[\alpha_1\beta_1- ga_1(\omega_1-gA_1)+g\alpha_1b_1]Q_1+ g^2a_1b_1P_1$,
$Z_1=[(\lambda_1-gB_1)P_1+ga_1\mu_0)]Q_1+ g^2a_1b_1P_1$, while for $j=2$ 
one should replace $a_1,A_1,b_1,B_1\to -(a_1+a_2),-(A_1+A_2),-(b_1+b_2),
-(B_1+B_2)$ and for $j=3$ one should substitute $a_1,A_1,b_1,B_1$ 
for $a_2,A_2,b_2,B_2$ to obtain the corresponding quantities 
$P_j, Q_j, Y_j, Z_j$, so that, 
for instance, $\beta_2=\sqrt{\mu_0^2-[\omega_2+g(A_1+A_2)]^2+g^2(b_1+b_2)^2}$,
$\beta_3=\sqrt{\mu_0^2-(\omega_3-gA_2)^2+g^2b_2^2}$. Also it should be noted 
that the quantum numbers $n_j$ of (9)--(11) are defined by the relations (for 
more details see Ref. \cite{Gon01})
$$n_1=\frac{gb_1Z_1-\beta_1Y_1}{\beta_1P_1Q_1}, 
n_2=-\frac{g(b_1+b_2)Z_2+\beta_2Y_2}{\beta_2P_2Q_2},
n_3=\frac{gb_2Z_3-\beta_3Y_3}{\beta_3P_3Q_3}.\>
\eqno(15)$$
Finally, $C_j$ is determined
from the normalization condition
$$\int_0^\infty(|F_{j1}|^2+|F_{j2}|^2)dr=\frac{1}{3}\>.\eqno(16)$$
Consequently, we shall gain that
$\Psi_j\in L_2^{4}({\Bbb R}^3)$ at any $t\in{\Bbb R}$ and, as a result,
the solutions of (7) may describe relativistic bound states of quarkonium
with the energy spectrum (9)--(11).

Now we can adduce numerical results for constants parametrizing
the charmonium spectrum which are shown in Table 1.

\begin{table}
\caption{Gauge coupling constant, mass parameter $\mu_0$ and
parameters of the confining SU(3)-connection for charmonium.}
\label{t.1}
\begin{center}
\begin{tabular}{cccccccc}
\noalign{\hrule}\\
$g$ & $\mu_0$ & $a_1$  & $a_2$ & $b_1$ & $b_2$ & $B_1$ &
$B_2$ \\
  & (GeV) &  &  & (GeV) & (GeV) & &  \\
\noalign{\hrule}\\
0.46900 & 0.62500 & 2.21104 & -0.751317 & 20.2395 & -12.6317 &
6.89659 &  6.89659 \\
\noalign{\hrule}\\
\end{tabular}
\end{center}
\end{table}
One can note that the obtained mass parameter $\mu_0$ is consistent with the
present-day experimental limits \cite{pdg} where the current mass of $c$-quark
($2\mu_0$) is accepted between 1.1 GeV and 1.4 GeV. As for the gauge coupling 
constant $g$ then its value has been chosen in accordance with many recent 
considerations \cite{gc}, wherefrom one can conclude that the strong coupling
constant $\alpha_s=g^2$ is of order $0.22\approx0.469^2$ at the scale of the 
$c$-quark current mass. 
At last, as to parameters $A_{1,2}$ of solution (3), it is clear that they only
shift the origin of count for the corresponding energies
and we can consider $A_1=A_2=0$.

With the constants of Table 1 the present-day levels of charmonium spectrum
were calculated with the help of (9)--(11) so Table 2 contains experimental 
values of these levels (from Ref. \cite{pdg})
and our theoretical ones computed according to the shown combinations
(we use the notations of levels from Ref. \cite{pdg}).

\begin{table}
\caption{Theoretical and experimental charmonium levels.}
\label{t.2}
\begin{center}
\begin{tabular}{c|c|c} 
\hline
State & Theoret. energy $\varepsilon_j=\sum\limits_{k=1}^3\omega_k$ &  
Experim. energy value \\
 & (GeV) & (GeV) \\
\hline
$\eta_c(1S)$ & $\varepsilon_1= \omega_1(0,0,-1)+\omega_2(0,0,-1)+
\omega_3(0,0,-1)= 2.979597$ & 2.979600 \\
\hline
$J/\psi(1S)$ & $\varepsilon_2= \omega_1(0,0,-1)+\omega_2(0,0,1)+
\omega_3(0,0,-1)= 3.096913$ & 3.096916 \\ 
\hline
$\chi_{c0}(1P)$ & $\varepsilon_3= \omega_1(0,0,-1)+\omega_2(0,0,-1)+
\omega_3(0,0,1)= 3.415186$  & 3.415190 \\
\hline
$\chi_{c1}(1P)$ &
$\varepsilon_4= \omega_1(0,0,1)+\omega_2(2,0,1)+\omega_3(0,1,-1)
= 3.505304 $ & 3.510590 \\
\hline
$h_{c}(1P)$ &
$\varepsilon_5= \omega_1(0,0,-1)+\omega_2(0,0,1)+\omega_3(0,0,1)
= 3.532503$ & 3.526210 \\
\hline
$\chi_{c2}(1P)$ &
$\varepsilon_6= \omega_1(0,1,-1)+\omega_2(1,1,-1)+\omega_3(1,1,-1)
= 3.553097$ & 3.556260 \\
\hline
$\eta_c(2S)$ &
$\varepsilon_7= \omega_1(0,0,1)+\omega_2(1,0,-1)+\omega_3(0,1,-1)
= 3.671608$ & 3.65400 \\
\hline
$\psi(2S)$ &
$\varepsilon_8= \omega_1(0,1,-1)+\omega_2(2,1,1)+\omega_3(1,1,-1)
= 3.674025$ & 3.685093 \\
\hline
$\psi(3770)$ &
$\varepsilon_9= \omega_1(0,0,1)+\omega_2(2,0,-1)+\omega_3(0,0,1)
= 3.775598$ & 3.770000\\
\hline
$\psi(3836)$ &
$\varepsilon_{10}= \omega_1(0,1,-1)+\omega_2(0,0,1)+\omega_3(0,1,1)
= 3.833640$ & 3.836000\\
\hline
$X(3872)$ &
$\varepsilon_{11}= \omega_1(0,1,-1)+\omega_2(0,1,1)+\omega_3(0,1,1)
= 3.871672$ & 3.872000\\
\hline
$\psi(4040)$ &
$\varepsilon_{12}= \omega_1(0,0,1)+\omega_2(1,1,1)+\omega_3(0,1,-1)
= 4.042660$ & 4.040000\\
\hline
$\psi(4160)$ &
$\varepsilon_{13}= \omega_1(0,0,-1)+\omega_2(0,0,-1)+\omega_3(0,1,1)
= 4.153765$ & 4.159000 \\
\hline
$\psi(4415)$ &
$\varepsilon_{14}= \omega_1(0,0,-1)+\omega_2(2,1,1)+\omega_3(1,0,-1)
= 4.409260$ & 4.415000 \\
\hline
\end{tabular}
\end{center}
\end{table}
\section{Specification of spectrum: electromagnetic transitions
$J/\psi(1S)\to \eta_c(1S)+\gamma$
and $\chi_{c0}(1P)\to J/\psi(1S)+\gamma$ in dipole approximation}

Now we should specify the obtained above charmonium spectrum. The fact is 
that the relations (9)--(11) permit various parametrizations of the charmonium
spectrum (see Refs. \cite{{GC03},{Gon03}}) and therefore it should impose 
further conditions to fix a certain parametrization among several possible 
ones. In the present paper we shall restrict ourselves to elecromagnetic 
transitions $J/\psi(1S)\to \eta_c(1S)+\gamma$ and 
$\chi_{c0}(1P)\to J/\psi(1S)+\gamma$ since we are interested in the first three
levels of charmonium. We shall compute widths of the mentioned transitions in 
dipole approximation that will allow us to use the corresponding wave functions 
written out in (13)--(14). Dipole approximation is often employed in meson 
physics, at least, as a primordial width estimate for one or another 
electromagnetic transition (see, e. g. Ref. \cite{BSG}) so we shall, without 
going into details, remind that in dipole approximation a width $\Gamma$ of an 
electromagnetic transition with emission of one photon for some system is 
given by the relation (for more 
details see, e. g. Ref. \cite{LL1})
$$\Gamma=\frac{4\omega^3}{3}|d_{fi}|^2\>,    \eqno(17)$$
where $d_{fi}$ -- a matrix element of the system dipole moment ${\bf d}=
Q(x{\bf i}+y{\bf j}+z{\bf k})$, $Q$ is an electromagnetic charge of system, 
the matrix element is taken between a final system stationary state and 
an initial one and
$\omega$ is the transition frequency. Let us introduce the quantities
$\xi=x+iy=r\sin{\vartheta}e^{i\varphi}, 
\eta=x-iy=r\sin{\vartheta}e^{-i\varphi}, z=r\cos{\vartheta}$, then it is clear
that
$$|d_{fi}|^2=Q^2\left[\frac{|\xi_{fi}|^2+|\eta_{fi}|^2}{2}+|z_{fi}|^2\right]\>,
\eqno(18)$$
while in our case $Q^2=(2e/3)^2$ with $e^2=1/137.0359895$. We should compute
the experimental widths $\Gamma(J/\psi(1S)\to \eta_c(1S)+\gamma)=
\Gamma_{94}=1.1557\ {\rm keV}$ and $\Gamma(\chi_{c0}(1P)\to J/\psi(1S)+\gamma)=
\Gamma_{16}=120.36\ {\rm keV}$ (data and notations from \cite{pdg}) with the 
transition frequencies $\omega_{94}=0.117316\ {\rm GeV}$, 
$\omega_{16}=0.318274\ {\rm GeV}$.
\subsection{Calculation of $\Gamma_{94}$}
As is clear from Table 2, the given transition is conditioned by 
the corresponding one of the second colour component 
$\omega_2(0,0,1)\to\omega_2(0,0,-1)$. The initial stationary wave function 
according to (7) is
$$\psi_2^i=r^{-1}\pmatrix{F_{21}^i\Phi_2^i\cr\
F_{22}^i\sigma_1\Phi_2^i}\>\eqno(19)$$
with the 2D eigenspinor $\Phi_2^i$ of the
euclidean Dirac operator ${\cal D}_2$ on the unit sphere ${\Bbb S}^2$ 
conforming to the eigenvalue $\lambda=\lambda_2=1$ while 
the final stationary wave function is
$$\psi_2^f=r^{-1}\pmatrix{F_{21}^f\Phi_2^f\cr\
F_{22}^f\sigma_1\Phi_2^f}\>\eqno(20)$$
with the 2D eigenspinor $\Phi_2^f$ of ${\cal D}_2$  
conforming to the eigenvalue $\lambda=\lambda_2=-1$.
Further we have $F^i_{21}=C^i_2P^i_2r^{\alpha^i_2}e^{-\beta^i_2r}\left[1+ 
\frac{g(b_1+b_2)}{\beta^i_2}\right]$, 
$F^i_{22}=iC^i_2Q^i_2r^{\alpha^i_2}e^{-\beta^i_2r}\left[1- 
\frac{g(b_1+b_2)}{\beta^i_2}\right]$ with 
$P^i_2=-g(b_1+b_2)+\beta^i_2$, $Q^i_2=\mu_0+\omega^i_2$, 
$\beta^i_2=\sqrt{\mu_0^2-(\omega^i_2)^2+g^2(b_1+b_2)^2}$, 
$\alpha^i_2=\sqrt{[1+g(B_1+B_2)]^2-g^2(a_1+a_2)^2}$, 
$\omega^i_2=\omega_2(0,0,1)$ as well as
$F^f_{21}=C^f_2P^f_2r^{\alpha^f_2}e^{-\beta^f_2r}\left[1+ 
\frac{g(b_1+b_2)}{\beta^f_2}\right]$, 
$F^f_{22}=iC^f_2Q^f_2r^{\alpha^f_2}e^{-\beta^f_2r}\left[1- 
\frac{g(b_1+b_2)}{\beta^f_2}\right]$ with
$P^f_2=-g(b_1+b_2)+\beta^f_2$, $Q^f_2=\mu_0+\omega^f_2$, 
$\beta^f_2=\sqrt{\mu_0^2-(\omega^f_2)^2+g^2(b_1+b_2)^2}$, 
$\alpha^f_2=\sqrt{[-1+g(B_1+B_2)]^2-g^2(a_1+a_2)^2}$, 
$\omega^f_2=\omega_2(0,0,-1)$ and we took into account that according to (15) 
$Y_2/Z_2=-g(b_1+b_2)/\beta_2$ at $n_2=0$. Also for constant $C^i_2$ 
we obtain from (16) (with the help of formula \cite{PBM1}
$\int_0^\infty x^{\alpha-1}e^{-px}dx=\Gamma(\alpha)p^{-\alpha}$,
Re $\alpha,p >0$) 
$$(C^i_2)^2\left[\left(1+\frac{g(b_1+b_2)}{\beta^i_2}\right)^2
(P^i_2)^2+
\left(1-\frac{g(b_1+b_2)}{\beta^i_2}\right)^2(Q^i_2)^2\right]
\frac{\Gamma(2\alpha^i_2+1)}{(2\beta^i_2)^{(2\alpha^i_2+1)}}=
\frac{1}{3}\>,\eqno(21)$$ 
whereas the analogous relation for $C^f_2$ can be obtained from (21) by 
changing indices $i\to f$.

At last, as for spinors $\Phi_2^i$, $\Phi_2^f$, it should be noted that
the eigenvalues of Dirac operator ${\cal D}_2$ have the multiplicity 
$2(l_2+1)$ for each $\lambda_2=\pm(l_2+1)$, so the spinors 
$\Phi_2^i$, $\Phi_2^f$ ($l_2 =0$) can be taken 
in the forms (for more details see Refs. \cite{Gon99})
$$\Phi_2^{i1}=\frac{C}{2}\pmatrix{\cos{\frac{\vartheta}{2}}+
i\sin{\frac{\vartheta}{2}}\cr
\cos{\frac{\vartheta}{2}}-i\sin{\frac{\vartheta}{2}}\cr}e^{i\varphi/2}, 
\Phi_2^{i2}=\frac{C}{2}\pmatrix{\cos{\frac{\vartheta}{2}}+
i\sin{\frac{\vartheta}{2}}\cr
-\cos{\frac{\vartheta}{2}}+i\sin{\frac{\vartheta}{2}}\cr}e^{-i\varphi/2},$$
$$\Phi_2^{f1}=\frac{C}{2}\pmatrix{\cos{\frac{\vartheta}{2}}-
i\sin{\frac{\vartheta}{2}}\cr
\cos{\frac{\vartheta}{2}}+i\sin{\frac{\vartheta}{2}}\cr}e^{i\varphi/2}, 
\Phi_2^{f2}=\frac{C}{2}\pmatrix{-\cos{\frac{\vartheta}{2}}+
i\sin{\frac{\vartheta}{2}}\cr
\cos{\frac{\vartheta}{2}}+i\sin{\frac{\vartheta}{2}}\cr}e^{-i\varphi/2},
\eqno(22) $$
where the coefficient $C=\sqrt{1/(2\pi)}$.

Under the circumstances we shall 
have [(*) signifies complex conjugation]
$$\xi_{fi}=\int(\psi_2^f)^{\dag}\xi\psi_2^id^3x=\int\limits_{{\Bbb R}^3}
r(F^{*f}_{21}F^i_{21} +F^{*f}_{22}F^i_{22})[\sum(\Phi^{f}_2)^{\dag}\Phi^{i}_2]
\sin^2{\vartheta}e^{i\varphi}dr d\vartheta d\varphi\>,
\eqno(23)$$
where $\sum(\Phi^{f}_2)^{\dag}\Phi^{i}_2$ is the sum over all possible 
combinations of final and initial spinors. It is then not complicated to see 
that the only nonzero contribution comes from the combination
$(\Phi_2^{f1})^{\dag}\Phi_2^{i2}=e^{-i\varphi}(iC^2\sin{\vartheta})/2$ 
while the rest of 
combinations give zero when integrating over $\varphi$ in (23). Fulfiling 
the remaining integration in (23) over $\vartheta$ and $r$ yields
$$\xi_{fi}=i\frac{2}{3}C^i_2C^f_2
\frac{\Gamma(\alpha^f_2+\alpha^i_2+2)}{(\beta^f_2+\beta^i_2)^{
\alpha^f_2+\alpha^i_2+2}}\times$$
$$\left\{P^f_2P^i_2
\left[1+\frac{g(b_1+b_2)}{\beta^f_2}\right]
\left[1+\frac{g(b_1+b_2)}{\beta^i_2}\right]+
Q^f_2Q^i_2
\left[1-\frac{g(b_1+b_2)}{\beta^f_2}\right]
\left[1-\frac{g(b_1+b_2)}{\beta^i_2}\right]\right\}\>.\eqno(24)$$
The similar considerations show that $\eta_{fi}\ne0$ only for 
the spinor combination 
$(\Phi_2^{f2})^{\dag}\Phi_2^{i1}=-e^{i\varphi}(iC^2\sin{\vartheta})/2$ which 
entails
$\xi_{fi}=-\eta_{fi}$ while two spinor combinations giving the contributions
to $z_{fi}$ are 
$(\Phi_2^{f2})^{\dag}\Phi_2^{i2}=-(\Phi_2^{f1})^{\dag}\Phi_2^{i1}=
\cos{\vartheta}/(8\pi)$ which gives rise to $z_{fi}=0$. As a result, we obtain 
according to (18)
$$\Gamma_{94}=\frac{4\omega^3_{94}}{3}Q^2|\xi_{fi}|^2\>  \eqno(25)$$
with $\xi_{fi}$ of (24).
\subsection{Calculation of $\Gamma_{16}$}
In accordance with Table 2 the given transition is formed by two ones:
the photon absorption with frequency  
$\omega_I=\omega_2(0,0,1)-\omega_2(0,0,-1)=0.117316$\ GeV by the second colour 
component and the photon emission with frequency 
$\omega_{II}=\omega_3(0,0,1)-\omega_3(0,0,-1)=0.435591$\ GeV by 
the third colour component so the resulting frequency
$\omega_{16}=\omega_{II}-\omega_{I}=0.318275$\ GeV. Accordingly we can write
$\Gamma_{16}=\Gamma_{I}+\Gamma_{II}$. Now the considerations analogous to the
above ones show that again $\xi_{fi}=-\eta_{fi}$, $z_{fi}=0$ in both 
transitions and $\Gamma_{I}$ is obtained from (24)--(25) by replacing $i\to f$, 
$f\to i$, $\omega_{94}\to\omega_{I}$ while $\Gamma_{II}$ will be the result 
of changing indices $2\to3$ and $(b_1+b_2)\to-b_2$, $(a_1+a_2)\to-a_2$, 
$(B_1+B_2)\to-B_2$, $\omega_{94}\to\omega_{II}$ in (24)--(25).
\subsection{Numerical results}
Table 3 gives the results of evaluation for $\Gamma_{94}$ and $\Gamma_{16}$
according to the obtained relations and the corresponding experimental ones 
of Ref. \cite{pdg}.
\begin{table}
\caption{Widths of electromagnetic transitions in charmonium.}
\label{t.3}
\begin{center}
\begin{tabular}{c|c|c} 
\hline
$\Gamma$ & Theoret. value &  Experim. value \\
 & (keV) & (keV) \\
\hline
$\Gamma_{94}$ &  $1.02977 $ & $1.15570$ \\  
\hline
$\Gamma_{16}$ &  6.06961  & 120.360 \\
\hline
\end{tabular}
\end{center}
\end{table}
Table 3 supplies us with an additional justification for the choice of parameters 
of the SU(3)-confining gluon field adduced in Table 1 and allows us to conclude 
that dipole approximation is not enough for the second transition of the ones 
under consideration. The question now is what gluon concentrations are in the 
mentioned SU(3)-confining gluon field.
\section{Estimates of gluon concentrations}
 To obtain necessary estimates we shall use an analogy with classical 
electrodynamics where is well known (see e. g. \cite{LL}) that the notion of 
classical electromagnetic field (a photon condensate) generated by a charged 
particle is applicable only at distances $>>$ the Compton wavelength 
$\lambda_c=1/m$ for the given particle. If denoting $\lambda_B$ the de Broglie 
wavelength of the particle then $\lambda_B=1/p$ with the relativistic impulse
$p=mv/\sqrt{1-v^2}$ while $v$ is the velocity of the particle (as a 
result, $\lambda_c=\lambda_B$ at $v=1/\sqrt{2}$) so one can 
rewrite $\lambda_B=\lambda_c\sqrt{1-v^2}/v$ and it is clear that 
$\lambda_B\to0$ when $v\to1$ (ultrarelativistic case), i. e., the particle 
becomes more and more point-like one. Accordingly, one can 
conclude that in the latter case the notion of classical electromagnetic field 
generated by a charged ultrarelativistic particle is applicable at distances 
$>>\lambda_B$. Under the circumstances, if the ultrarelativistic charged 
particle accomplishes its motion within the region with characteristic 
size of order $r_0$ then in the given region the electromagnetic field 
generated by the particle may be considered as classical one at
$r_0>>\lambda_B$. For example, in the case of positronium we have $r_0\sim
2a_0$, where $a_0\approx\ 5.29\cdot10^{-11}\ {\rm m}$ is the Bohr radius, so
$r_0>>\lambda_e\approx\ 3.86\cdot10^{-13}\ {\rm m}$, the electron Compton 
wavelength, i. e., the electric Coulomb interaction between electron and
positron in positronium can be considered as classical electromagnetic field.
  
Passing on to QCD, gluons and quarkonia, it should be noted that quarks 
in quarkonia accomplish a finite motion within 
a region of order $1\ {\rm fm}=10^{-15}\ {\rm m}$. Then, as 
is seen from the radial parts of the wave functions (13)--(14), the quantity 
$1/\beta_j$ permits to be considered
as a characteristic size of the $j$-th colour component of the given quarkonium 
state and, consequently, we can take the magnitude 
$$r_0=\frac{1}{3}\sum_{j=1}^{3}\frac{1}{\beta_j}\>\eqno(26)$$
for a characteristic size of the whole quarkonium state and, in line with 
the above, we should consider the confining SU(3)-gluonic Yang-Mills field of
(3) or (6) to be classical one when $r_0>>\lambda_B$, the de Broglie 
wavelength of the corresponding quarks forming quarkonium.

  On the other hand, a classical electromagnetic field (photon condensate) 
conforms to the large photon concentrations for every frequency 
presented in the field \cite{LL1}. Then in QCD we should require the large 
gluon concentrations in the given classical gluonic field (gluon condensate).
To estimate the given concentrations
we can employ $T_{00}$-component of the energy-momentum tensor of (5)
and, taking the quantity 
$\omega= \Gamma$, the whole decay width of the quarkonium state, for 
the characteristic frequency we obtain
the sought characteristic concentration $n$ in the form
$$n=\frac{T_{00}}{\Gamma}\>. \eqno(27)$$

It is not complicated to obtain the curvature matrix (field strentgh) 
corresponding to the solution (3) or (6)
$$F= F^a_{\mu\nu}\lambda_a dx^\mu\wedge dx^\nu=
-\partial_r(A^a_t\lambda_a)dt\wedge dr
+\partial_r(A^a_\varphi \lambda_a)dr\wedge d\varphi
\>, \eqno(28)$$
which entails the only nonzero components
$$F^3_{tr}=-\frac{2a_1+a_2}{2r^2},\>
F^8_{tr}=\frac{a_2\sqrt{3}}{2r^2},\>
F^3_{r\varphi}=b_1+\frac{b_2}{2},\>
F^8_{r\varphi}=-\frac{b_2\sqrt{3}}{2}\>\eqno(29) $$
and, in its turn,
$$T_{00}\equiv T_{tt}=\frac{1}{4\pi}\left\{\frac{3}{4}\left[(F^3_{tr})^2+
(F^8_{tr})^2\right]
+\frac{1}{4r^2\sin^2{\vartheta}}\left[(F^3_{r\varphi})^2+
(F^8_{r\varphi})^2\right]\right\}= $$
$$\frac{3}{16\pi}\left(\frac{a_1^2+a_1a_2+a_2^2}{r^4}+
\frac{b_1^2+b_1b_2+b_2^2}{3r^2\sin^2{\vartheta}}\right),\>\eqno(30)$$
so, further putting $\sin^2{\vartheta}=1/3$ for simplicity, we can rewrite
(30) in the form
$$T_{00}=T_{00}^{\rm coul}+T_{00}^{\rm lin}\>\eqno(31)$$
conforming to the contributions from the Coulomb and linear parts of the
solution (3) or (6). The latter gives the corresponding split of $n$ from
(27)
$$n=n_{\rm coul} + n_{\rm lin}.\>\eqno(32)$$
For comparison we shall also estimate the photon concentration in the ground 
state of the positronium. As is known historically \cite{Per}, the analogy 
between the latter 
system and quarkonia played the important part when building the quarkonia 
models. For positronium we have the electromagnetic Coulomb interaction
$A=A_tdt=(e/r)dt$ which entails $F=F_{tr}dt\wedge dr=(e/r^2)dt\wedge dr$ and
$$T_{00}\equiv T_{tt}=\frac{1}{4\pi}(F_{tr})^2=\frac{\alpha_{em}}{4\pi r^4}
\>\eqno(33)$$
with $\alpha_{em}=e^2=1/137.0359895$.

\section{Numerical results}
When computing for the first three states of charmonium 
we used their present-day whole decay widths 
$\Gamma=17.3\ {\rm MeV}$, $91.0\ {\rm keV}$, $10.2\ {\rm MeV}$ 
respectively \cite{pdg},
while the calculation $r_0$ of (26) gives
$r_0=r_1, r_2, r_3$ (see Table 4).

In the positronium case we employed the widths $\Gamma_0=1/\tau_0$ 
(parapositronium) and $\Gamma_1=1/\tau_1$ (orthopositronium), respectively, 
with the life times $\tau_0=1.252\cdot10^{-10}\ {\rm s}$, 
$\tau_1=1.377\cdot10^{-7}\ {\rm s}$ \cite{Per} while $r_0=2a_0$ with the
Bohr radius $a_0=0.529177249\cdot10^{5}\ {\rm fm}$ \cite{pdg}.

Tables 4, 5 contain the numerical results for both the cases, where we used
$1\ {\rm \AA}=10^{-10}\ {\rm m}=10^{5}\ {\rm fm}$. Finally, when calculating 
we applied the relation 
$1\ {\rm GeV^{-1}}\approx0.21030893\ {\rm fm}\>$.
\begin{table}
\caption{Gluon concentrations in the first three states of charmonium.}
\label{t.4}
\begin{center}
\begin{tabular}{cccc}
\noalign{\hrule}\\
$\eta_c(1S)$: & $r_1= 0.0399766\ {\rm fm}$ &  & \\
\noalign{\hrule}\\
$r$ &  $n_{\rm coul}$ & $n_{\rm lin}$ & $n$  \\
fm & $ ({\rm fm}^{-3}) $ & (${\rm fm}^{-3}) $ & (${\rm fm}^{-3}) $ \\
\noalign{\hrule}\\
$0.1r_1$ & $0.550649\cdot10^{13}$   & $0.727630\cdot10^{10}$ & 
$0.551377\cdot10^{13}$  \\
$r_1$ & $0.550649\cdot10^{9}$ & $0.727630\cdot10^8$ & $0.623412\cdot10^{9}$  \\
$10r_1$ & $0.550649\cdot10^5$  & $0.727630\cdot10^6$ & $0.782695\cdot10^6$  \\
$1.0$ & $0.140637\cdot10^{4}$  & $0.116285\cdot10^{6}$ & 
$0.117691\cdot10^{6}$  \\
$a_0$ & $0.179347\cdot10^{-15}$  & $0.415260\cdot10^{-4}$ & 
$0.415260\cdot10^{-4}$  \\
\noalign{\hrule}\\
$J/\psi(1S)$: & $r_2= 0.0397797\ {\rm fm}$ &  & \\
\noalign{\hrule}\\
$r$ &  $n_{\rm coul}$ & $n_{\rm lin}$ & $n$  \\
fm & $ ({\rm fm}^{-3}) $ & (${\rm fm}^{-3}) $ & (${\rm fm}^{-3}) $ \\
\noalign{\hrule}\\
$0.1r_2$ & $0.106772\cdot10^{16}$   & $0.139702\cdot10^{13}$ & 
$0.106912\cdot10^{16}$  \\
$r_2$ & $0.106772\cdot10^{12}$ & $0.139702\cdot10^{11}$ & 
$0.120742\cdot10^{12}$  \\
$10r_2$ & $0.106772\cdot10^8$  & $0.139702\cdot10^9$ & $0.150380\cdot10^9$  \\
$1.0$ & $0.267364\cdot10^{6}$  & $0.221069\cdot10^{8}$ & 
$0.223742\cdot10^{8}$  \\
$a_0$ & $0.340956\cdot10^{-13}$  & $0.789450\cdot10^{-2}$ & 
$0.789450\cdot10^{-2}$  \\
\noalign{\hrule}\\
$\chi_{c0}(1P)$: & $r_3= 0.0401851\ {\rm fm}$ &  & \\
\noalign{\hrule}\\
$r$ &  $n_{\rm coul}$ & $n_{\rm lin}$ & $n$  \\
fm & $ ({\rm fm}^{-3}) $ & (${\rm fm}^{-3}) $ & (${\rm fm}^{-3}) $ \\
\noalign{\hrule}\\
$0.1r_3$ & $0.914712\cdot10^{13}$   & $0.122134\cdot10^{11}$ & 
$0.915933\cdot10^{13}$  \\
$r_3$ & $0.914712\cdot10^{9}$ & $0.122134\cdot10^{9}$ & 
$0.103685\cdot10^{10}$  \\
$10r_3$ & $0.914712\cdot10^5$  & $0.122134\cdot10^7$ & 
$0.131282\cdot10^7$  \\
$1.0$ & $0.238531\cdot10^{4}$  & $0.197228\cdot10^{6}$ & 
$0.199613\cdot10^{6}$  \\
$a_0$ & $0.304186\cdot10^{-15}$  & $0.704313\cdot10^{-4}$ & 
$0.704313\cdot10^{-4}$  \\
\noalign{\hrule}\\
\end{tabular}
\end{center}
\end{table}

\begin{table}
\caption{Photon concentrations in the ground state of positronium.} 
\label{t.5}
\begin{center}
\begin{tabular}{ccc}
\noalign{\hrule}\\
   & $r_0= 2a_0=2\cdot0.529177249\cdot10^{5}$\ {\rm fm} &  \\
\noalign{\hrule}\\
   & Parapositronium & Orthopositronium   \\
$r$ &  $n_{\rm para}$ & $n_{\rm ortho}$   \\
fm & (${\rm fm}^{-3}$ or ${\rm \AA}^{-3}$) &
 (${\rm fm}^{-3}$ or ${\rm \AA}^{-3}$)  \\
\noalign{\hrule}\\
$0.01r_0$ & $0.888025\cdot10$ ($0.888025\cdot10^{16}$) & 
$0.976685\cdot10^4$ ($0.976685\cdot10^{19}$)  \\
$0.1r_0$ & $0.888025\cdot10^{-3}$ ($0.888025\cdot10^{12}$)  & 
0.976685 ($0.976685\cdot10^{15}$) \\
$r_0$ & $0.888025\cdot10^{-7}$ ($0.888025\cdot10^{8}$)  & 
$0.976685\cdot10^{-4}$ ($0.976685\cdot10^{11}$)   \\
$2r_0$ & $0.555015\cdot10^{-8}$ ($0.555015\cdot10^{7}$) & 
$0.610428\cdot10^{-5}$ ($0.610428\cdot10^{10}$)   \\
\noalign{\hrule}\\
\end{tabular}
\end{center}
\end{table}

\section{Concluding remarks}
It is clear that for charmonium the natural unit of the gluon concentration is
${\rm fm^{-3}}$ while for positronium the photon concentration should be 
measured in ${\rm \AA}^{-3}$. Then, as is seen from Tables 4, 5, qualitative 
behaviour of both the concentrations is similar. At the characteristic scales
of each system the concentrations are large and the corresponding fields 
(electric and magnetic colour ones or electric Coulomb one) can be considered
as classical ones. For charmonium the part $n_{\rm coul}$ of gluon concentration 
$n$ connected with the Coulomb colour electric field is decreasing faster than 
$n_{\rm lin}$, 
the part of $n$ related to the linear colour magnetic field, and at large 
distances $n_{\rm lin}$ becomes dominant. Under the circumstances, as has been
said in Section 4, we can estimate the quark velocities in the charmonium 
states under discussion from the condition 
$$v=\frac{1}{\sqrt{1+\left(\frac{\lambda_B}{\lambda_q}\right)^2}}\>
\eqno(34)$$
with the $c$-quark Compton wavelength 
$\lambda_q=1/(2\mu_0)\approx0.168247\ {\rm fm}$ so, taking the de Broglie 
wavelength $\lambda_B=0.1r_k$ ($k=1, 2, 3$) with $r_k$ from Table 4,
we obtain $v_1\approx0.999718$, $v_2\approx0.999721$, $v_3\approx0.999715$ 
i. e., the quarks in
charmonium should be considered as the ultrarelativistic point-like particles.
This additionally confirms the conclusion of Refs. \cite{{GC03},{Gon03}} that
the relativistic effects are extremely important for the confinement mechanism.
As a result, the confinement scenario described early in Section 1 may really
occur.

The results of the given paper can be extended to both other states of 
charmonium and the bottomonium states. We hope to return to the problem
elsewhere.


\end{document}